\documentstyle[12pt,amssym,aasms4]{article}

\received{}
\revised{}
\accepted{}
\slugcomment{(Accepted for publication in ApJ Letters)}

\lefthead{Sahu \& Blades}
\righthead{UV absorption lines towards NGC~1705}

\newcommand{\vlsr}{$v_{\mbox{\tiny LSR}}$}

\newcommand{\kms}{$\mbox{km\,s}^{-1}$}

\begin{document}

\title{UV Interstellar Absorption Lines towards the Starburst Dwarf Galaxy 
NGC~1705 \footnotemark}

\author{M. S. Sahu and J. Chris Blades}
\affil{Space Telescope Science Institute, 3700 San Martin Drive, Baltimore, MD~21218, msahu@stsci.edu, blades@stsci.edu}


\footnotetext{Based on observations taken with the NASA/ESA $\it{Hubble~
Space~Telescope}$, obtained at the Space Telescope Science Institute, which
is operated by the Association of the Universities for Research in Astronomy
under contract NAS 5-26555}

\begin{abstract}
Archival Goddard High Resolution Spectrograph low-resolution spectra  of NGC~1705, with wavelength ranges 1170.3 to 1461.7 \AA \ and 1453.5 to 1740.1
\AA \ and a velocity resolution
$\sim$ 100 \kms, have been used to derive the velocity structure and equivalent widths of the absorption lines of Si~{\sc ii} $\lambda$1190.42, 
$\lambda$1260.42, $\lambda$1304.37 and $\lambda$1526.71 \AA, S~{\sc ii} $\lambda$1253 \AA, Al~{\sc ii} $\lambda$1670.79 \AA \ and Fe~{\sc ii} $\lambda$1608.45 \AA \ in this sightline. Three relatively narrow absorption components are seen at LSR velocities --20 \kms\ , 260 \kms\ and 540 \kms\ . Arguments are presented to show these absorption features are interstellar rather than stellar in origin based on a comparison with the 
C~{\sc iii} $\lambda$1175.7 \AA \ absorption feature. 
We identify the --20 \kms\ component with  Milky Way disk/halo gas and the 260 \kms\ component with an isolated high-velocity cloud HVC~487. This small HVC is located $\sim$ 10$^\circ$ from the H~{\sc i} gas which envelops the Magellanic Clouds and the Magellanic Stream (MS). The (Si/H) ratio for this HVC is $>$ 0.6 (Si/H)$_\odot$ which together with velocity agreement, suggests association with the Magellanic Cloud and MS gas.

H$_{\alpha}$ emission line kinematics of NGC~1705 show the presence of a kpc-scale expanding supershell of ionized gas centered on the central nucleus with a blue-shifted emission component at 540 \kms\  (Meurer et al. 1992).  We identify the 540 \kms\ absorption component seen in the $GHRS$ spectra with the front side of this expanding, ionized supershell. 
The most striking feature of this component is strong Si~{\sc ii} and Al~{\sc ii}  absorption but weak  Fe~{\sc ii} $\lambda$1608 \AA \ absorption. The low 
Fe~{\sc ii} column density derived is most likely intrinsic since it cannot be accounted for by ionization corrections or dust depletion. 

Due to their shallow gravitational potential wells, dwarf galaxies have small
gravitational binding energies and are vulnerable to large mass losses from strong winds driven by the supernovae from the first generation of stars. Galactic winds from dwarf galaxies occur at time-scales $<$ 10$^{8}$ years, which is less than the time-scale required to produce Type Ia supernovae. Consistent with our observations, shells produced by galactic winds are  expected to be enriched with Type II supernova products like Si, Al, Mg  and should be deficient in the products of Type Ia supernovae, like Fe and iron-peak elements. 
\end{abstract}

\keywords{ISM: abundances --- ultraviolet: ISM --- galaxies: ISM --- galaxies: starburst --- galaxies: individual (NGC~1705) --- galaxies: evolution}


%

\section{Introduction}

 NGC~1705 is a nearby, H~{\sc i}-rich dwarf galaxy with a systemic heliocentric velocity of 628 $\pm$ 9 \kms\ derived from the peak of the H~{\sc i} profile, corresponding to \vlsr\  = 610 \kms\ (Meurer et al. 1992, hereafter referred to as MFDC). H$_\alpha$ continuum subtracted images are striking: NGC~1705 is dominated by extended loops of H$_\alpha$ emission (see Fig. 2, MFDC). The continuum image shows a major axis extent of $\sim$ 40 arcsec and a minor axis extent of 20 arcsec, while the H$_\alpha$ loops extend beyond 40 arcsec  in the direction perpendicular to the minor axis. At least eight major loops and arcs have been catalogued from the H$_\alpha$ images by MFDC, and are reminiscent of  H~{\sc i} supershell structures seen in our Galaxy (Heiles, 1990). A super-star cluster NGC~1705A, with a mean spectral type B3 V, is present in the center of the galaxy (Melnick et al. 1985). The presence of the extended H~{\sc i} loops and the central super-star cluster point to the existence of a supernova-driven galactic wind in NGC~1705.

Previous studies with the $IUE$ revealed the presence of prominent absorption lines in the ultraviolet spectrum of NGC~1705 which York et al. (1990) argued were of an interstellar origin. However the data were at the sensitivity limit of $IUE$, and so we have used better quality spectra in the HST archive to 
re-investigate their nature. As we shall show, the features are indeed
(predominantly) interstellar in origin, and can be identified with three 
separate environments, namely, gas in our Galaxy, in NGC~1705 itself as well
as in a probable Magellanic Stream cloud near the sightline. The large velocity
range inherent in the absorption lines explains why they were so prominent in the lower resolution $IUE$ spectra. Discovery of the Magellanic Stream component
was unexpected. However, our most intriguing result concerns the component attributed to NGC~1705 which shows an underabundance of Fe compared to Si, S 
and Al. We discuss this underabundance within the context of supernova-driven galactic wind evolution of dwarf galaxies.

\section{The data}

Two $GHRS$ spectra of NGC~1705 taken through the small science aperture with the G140L grating were retrieved from the HST Data Archives : the first extending from 1170.3 to 1461.7 \AA \ and the second from 1453.5 to 
1740.1 \AA. The observations were made on 12 November, 1994 and obtained with a comb-addition of 4 to reduce detector fixed-pattern noise. The shorter wavelength spectrum has an exposure time of 1196.8 secs (filename Z2JT0107T) while the other spectrum (filename Z2JT0109T) has an exposure time of 2393.6 secs. The spectra were reduced using the standard pipeline software available in the STSDAS HRS package together with the the most recent calibration files. Individual WAVECAL exposures were taken before the science exposures and, since no FP-SPLIT was used, the ZWAVCAL task was used together with the WAVEOFF task to give an average wavelength offset value and improve the wavelength calibrations. For the G140L grating the dispersion is 0.57 \AA \
per diode, corresponding to a velocity resolution of 
 $\sim$ 140 \kms\ at $\lambda$1200 \AA \  and $\sim$ 100 \kms\ at $\lambda$1700 \AA \ . The wavelength accuracy is around $\sim$ 40 \kms\  . The signal-to-noise ratio (S/N) in the continuum ranges from $\sim$ 7:1
at 1670 \AA \  to 16:1 at 1400 \AA \ . There is good velocity agreement between the galactic component of the Si~{\sc ii} $\lambda$1190 \AA \
 and $\lambda$1526 \AA \  lines present in separate spectra, suggesting that the relative velocity scales of the two G140L spectra are consistent. The continuum is smooth and slowly varying and spectra were normalized by fitting low-order
polynomial functions.

Both spectra show complex, multi-component absorptions from all the common interstellar species. Column 1, 2, 3, 4 and 5 of Table 1 contain the ion name, vacuum wavelength, oscillator strength, measured equivalent widths and their uncertainties and the S/N in the continuum respectively. The equivalent width uncertainties quoted, include contributions from the continuum-placement uncertainties and noise. We have performed an iterative, non-linear least-square
fit to the observed profiles (see Blades et al. 1997). Figure 1 shows the continuum-normalized profiles of the Si~{\sc ii}, C~{\sc iii}, S~{\sc ii}, Fe~{\sc ii} and Al~{\sc ii} absorption lines plotted as histograms against LSR velocity, with model fits plotted as continuous lines. 
Other species are seen in the spectra, including C~{\sc i}, C~{\sc i}$^*$, O~{\sc i}, Si~{\sc iv} and C~{\sc iv}. A detailed line list and atlas of UV absorption lines for the NGC~1705 sightline will be published elsewhere.

\section{Discussion}
\subsection{Interstellar origin of the absorption lines}

York et al. (1990) considered the prominent absorption they found in the
NGC~1705 was interstellar. The superior HST data are compelling on this point.
First of all, we note that the C~{\sc iii} $\lambda$1175.7 \AA \ absorption feature is broad (EW = $\sim$ 1.8 \AA \ ), and centred at $\sim$ 620 \kms \ , close to the LSR recession velocity of NGC~1705. This C~{\sc iii} absorption feature, commonly seen in spectra of OB stars is a photospheric line and originates from the massive stars in the central superstar cluster NGC~1705A.

The Si~{\sc ii} $\lambda$1526.71 \AA \ absorption is unaccompanied by Si~{\sc ii} $\lambda$1533 \AA \ and must be interstellar (Walborn et al., 1995). This line has three narrow components :  a strong feature centred at --20 \kms \ , a relatively weak feature centred at  260 \kms\ , and a strong feature centred at  540 \kms\ . The other lines in Figure 1 show similar velocity structure. Their
velocities are not consistent with the systemic velcoity of NGC~1705, nor do any of the profiles show asymmetric or P-Cygni shapes characteristic of stellar mass loss. Hence, apart from the C~{\sc iii} feature, we can attribute the other lines to an interstellar origin without ambiguity. In the next section we
discuss where the individual components arise.

\subsection{Curve-of-growth analysis}

We have performed a curve-of-growth (cog) analysis for the three components at --20 \kms\ ,  260 \kms\ and 540 \kms\ and obtain best-fitting $b$-values of $\sim$ 25 \kms\ , $\sim$ 50 \kms\ and $\sim$ 30 \kms\  respectively. Details of the cog analysis will be presented elsewhere. Only lines with  optical depth at the centre of the line , $\tau$$_\circ$ $\leq$ 1 in Table 1 were used for column density determinations ; stronger lines provide lower limits. We note that the three components have large $b$ values compared to single, cold components seen in the ISM and must be due to successive overlapping components.
In our analysis we are interested only in the contribution to the column density from the large $b$-value components.

\subsection {Origin of the absorption features}

We attribute the absorption feature centred at --20 \kms \ to Milky Way disk and halo gas. This component shows strong Si~{\sc ii}, Fe~{\sc ii} and Al~{\sc ii} absorption and relatively weak S~{\sc ii} absorption. 

\subsubsection{The 260 \kms\ feature}

The NGC~1705 sightline intercepts a region which is approximately 10$^\circ$ away from the outermost H~{\sc i} contour that envelops both the Large and Small Magellanic Clouds (Mathewson \& Ford, 1984). We checked the southern 21-cm survey of high-velocity gas between --650 and 650 \kms\ by Bajaja et al. (1985) for H~{\sc i} gas at velocities corresponding to any of the UV absorption 
components. These authors reported the detection of an isolated cloud HVC~487 at (l,b) = (258.$^\circ$03, --39.$^\circ$15) and \vlsr\ = 232 \kms\ which is 
only $\sim$ 2$^\circ$ away from the sightline to NGC~1705. We attribute the 260 \kms\ absorption component  to the isolated HVC. Since this H~{\sc i} survey has a grid 2 $\times$ 2$^\circ$, and a beam size of 34$^\prime$, it is not possible to determine the true extent of this cloud. The velocity difference between the UV and H~{\sc i} feature is not significant because of the following reasons: (1) the H~{\sc i} data have a beam size of 34$^\prime$ while
the GHRS spectra has a beam with an infinitesimally small solid angle; (2) the
H~{\sc i} observations were located slightly away from the NGC~1705 sightline;
and (3) the velocity resolution of the GHRS data is $\sim$ 100 \kms\ with a centroid precision of 40 \kms\ while the H~{\sc i} data has a resolution of 16 \kms\ . 

The N(H~{\sc i}) value at the position of NGC~1705 is $\leq$ 2$\times$ 10$^{18}$ cm$^{-2}$  (Wakker \& van Woerden, 1991) and we find N(Si~{\sc ii}) = 4 $\times$ 10$^{13}$ cm$^{-2}$, implying  (Si/H) is $>$ 0.6 (Si/H)$_\odot$, consistent with the metal abundance estimates for the Magellanic Clouds and the Magellanic Stream (Blades et al. 1988, Lu et al. 1994) and together with velocity agreement, suggest close association. 
 
\subsubsection {The 540 \kms\ feature}

NGC~1705 is known to have strong outflows of gas: MFDC found the H$\alpha$ and [OIII] 5007 \AA \ nebular emission lines to be double-peaked with a separation of $\sim$ 100 \kms\ over most of the face of the galaxy. The kinematics of the ionized gas is best represented by a homogeneously expanding ellipsoidal shell
and is plotted in Fig 14 of MFDC. The blue shifted emission component of this kpc-scale expanding supershell is at  540 \kms\ , the \vlsr\ velocity of our UV absorption feature.

The most interesting property of this 540 \kms\ supershell component is strong Si~{\sc ii} and  Al~{\sc ii} absorption but relatively weak Fe~{\sc ii} absorption. We find N(Si~{\sc ii}) $\geq$  7.9 $\times$ 10$^{14}$ cm$^{-2}$, N(S~{\sc ii}) = 2.5 $\times$ 10$^{14}$ cm$^{-2}$, N(Fe~{\sc ii}) = 2.5 $\times$ 10$^{14}$ cm$^{-2}$ and N(Al~{\sc ii}) $\geq$ 1.5 $\times$ 10$^{14}$ cm$^{-2}$. The (Si/S) ratio for this component is $\geq$ 0.4 (Si/S)$_\odot$ suggesting low dust depletion in this supershell component.  For the 540 \kms\ component, ionization corrections cannot explain the observed low Fe~{\sc ii} column density because both Al and Fe have similar ionization potentials for the ionization X$^{2+}$ $\rightarrow$ X$^{3+}$ (18.83 and
16.16 eV respectively) and both Al and Fe have small cross sections for this ionization stage. Further, using the mean spectral type of B3V (Melnick et al., 1985) for the central superstar cluster NGC~1705A, the photoionization code CLOUDY (Ferland, 1993) and a plane parallel slab cloud with solar metallicity, we estimate Al~II/Al~III $>$ 500 and Fe~II/Fe~III $>$ 50. We conclude that the presence of gas in unobserved stages of ionization is not a significant factor. Secondly, dust depletion can not explain the low Fe abundance. Al and Fe have similar condensation temperatures (Jenkins, 1987)
similar first ionization potentials, 5.99 and 7.87 eV respectively and similar logarithmic gas-phase depletion in the diffuse ISM,  with D(Fe) and D(Al) ranging between -3.0 to -1.0 (Barker et al. 1984, Jenkins, 1987 and Savage \& Sembach, 1996). Moreover, within a given cloud with an overall level of depletion, the change in element-to-element depletion is 
$<$ 0.16 dex and significantly less than the element-to-element variation (Joseph, 1988 and Welty et al. 1997). The logarithmic depletion factor (D(X) = log (X/S) - log(X/S)$_{\odot}$) for the supershell component is D(Fe) = -0.87 which should be equal to D(Al)$\pm$0.16, making the $\tau$$_\circ$ for the Al~{\sc ii} line significantly lower than what is observed (Table 1). The low Fe~{\sc ii} column density for the 540 \kms\ cloud must therefore be intrinsic. We have illustrated this point further by using the N(S~{\sc ii}) for the 540 \kms\ component and solar (Fe/S) ratio, 
to estimate the column density $N$ for the same $b$ value.  This estimated profile is plotted in long dashed lines in Fig 1 over the Fe II $\lambda$1608.45 \AA\ profile. The column density estimated in this manner is a factor $\sim$ 10 greater than the observed Fe column density. Unfortunately the $\lambda$1608 \AA\ line is the only Fe line in the spectra and it will be extremely important to re-observe this sightline to
confirm if the other Fe lines show this underabundance.

\subsection{Gas expelled from NGC~1705 through supernova-driven galactic winds --- an overabundance of $\alpha$-process elements ?}

Could overabundance of Al, Si and S and $\alpha$-process elements with respect to iron-peak elements (Fe, Cr and Zn) in the gas expelled from a dwarf galaxy by a supernova driven galactic-wind explain the weak Fe~{\sc ii} line? Larson (1974) has argued that dwarf galaxies, because of their shallow gravitational potential wells, have small binding energies and are vulnerable to huge mass losses through winds driven by supernova explosions from the first generation of star formation. Dekel and Silk (1986) have shown that galaxies with velocity dispersions $<$ 100 \kms\ would lose much of their gas due to strong galactic winds. Galactic winds from dwarf galaxies (with total mass $\le$ 10$^{11}$ M$_{\odot}$) occur at time-scales $<$ 10$^8$ years (Nath and Chiba, 1995) which is less than the time required to produce Type Ia supernovae (Wheeler et al., 1989). Type Ia SNe, which produce nearly 2/3 of the iron and iron-peak elements (Truran and Bunkert, 1993) cannot contribute to the metallicity of this gas which has been produced by the first generation of stars and expelled from the dwarf galaxy. Shells produced by galactic winds are therefore expected to be enriched with Type II supernova products like Si and Al and deficient in Fe and other iron-peak elements which are produced mainly by Type Ia supernovae. The underabundance of Fe for the supershell component in the $GHRS$ spectra, is consistent with this interpretation. 

In addition, the 540 \kms \ absorption component in the UV spectra  
provides evidence that superwinds from dwarf galaxies could account for some types of QSO absorption-lines as suggested by York et al. (1986). Further HST-STIS observations of other strong iron and iron-peak element lines and $\alpha$-elements in wavelength ranges $\lambda$1750 to $\lambda$3000 \AA \ (not covered by our $GHRS$ spectra) would be extremely useful in confirming this result.

\acknowledgements

This work was funded through HST grants GO-3761 and GO-6723. We thank Don York for useful discussions, Bart Wakker for the 21-cm data and Nolan Walborn and Ralph Bohlin for their comments. We also thank an anonymous referee for helpful comments.

\clearpage
 
\begin{deluxetable}{cclccccccc}
\footnotesize
\tablecaption{Equivalent width measurements \label{tbl-1}}
\tablewidth{0pt}
\tablehead{
\colhead{Species} & \colhead{$\lambda$$_{vac}$} & \colhead{$f$}  & \colhead{} &
\colhead{EW (m\AA \ )} & \colhead{} &
\colhead{$S/N$} & \colhead {} & \colhead{$\tau$\tablenotemark{a}$_{\circ}$} & \colhead{} \\
\colhead{} & \colhead{(\AA \ )} & \colhead{} & 
\colhead{--20 \kms\ } & 
\colhead{260 \kms\ } & \colhead{540 \kms\ }  & \colhead {} & \colhead{--20} & \colhead{260} & \colhead{540}
} 
\startdata
Si~{\sc ii} & 1190.42  & 0.2502\tablenotemark{b} &  $449^{+26}_{-22}$ & $88^{+10}_{-12}$ & $\it{blend}$ & 16.0 &3.9 &0.25 & \nodata \nl
Si~{\sc ii} & 1260.42  & 1.007\tablenotemark{b} &  $\it{blend}$     & $\it{blend}$    & $704^{+14}_{-22}$ & 13.2 & \nodata & \nodata & 83 \nl
Si~{\sc ii} & 1304.37  & 0.086\tablenotemark{b} & $\it{blend}$  & $157^{+47}_{-34}$ & $370^{+27}_{-19}$ & 11.8 & \nodata & 0.42 & 1.85 \nl
Si~{\sc ii} & 1526.71  & 0.110\tablenotemark{c} & $399^{+31}_{-30}$ & $239^{+62}_{-28}$ & $661^{+21}_{-19}$ & 14.6 & 5.5 & 0.4 & 8.6 \nl
S~{\sc ii} & 1253.81 & 0.01088\tablenotemark{b}& $151^{+66}_{-18}$ & \nodata & $181^{+86}_{-56}$ & 12.8 & 0.4 & \nodata & 0.26 \nl 
Fe~{\sc ii} & 1608.45 & 0.062\tablenotemark{b}& $314^{+63}_{-20}$ & \nodata & $286^{+15}_{-44}$ & 9.6 & 2.3 & \nodata & 0.5 \nl
Al~{\sc ii} & 1670.79 & 1.88\tablenotemark{b} & $508^{+70}_{-81}$ & \nodata & $624^{+45}_{-20}$ & 6.9 & 6.4 & \nodata &  11.4 \nl
\nl
\enddata


\tablenotetext{a} {optical depth at the center of the line for the --20, 260 and 540 \kms\ components} 
\tablenotetext{b} {Morton, 1991}
\tablenotetext{c} {Spitzer \& Fitzpatrick, 1993}

\end{deluxetable}

\clearpage

%
%

\figcaption[fig1.epsi]{Normalized profiles plotted against LSR velocity are shown as histograms whereas best-fit model absorption line profiles are shown as continuous thin lines. The three absorption components at --10 \kms\ , 260 \kms\ and 540 \kms\ are shown with arrows on the Si~{\sc ii} $\lambda$1526.71 \AA \ profile. Comparison of this Si~{\sc ii} profile with the broad stellar C~{\sc iii}  profile provides evidence that the three absorption components are interstellar in origin. Note the strong absorption at 540 \kms\ in the Si~{\sc ii}, S~{\sc ii} and 
Al~{\sc ii} lines. The Fe~{\sc ii} absorption at this velocity, on the hand, is relatively weak. Ionization corrections and dust depletion ($\S$ 3.3.2) cannot account for the low Fe column density derived which must therefore be intrinsic. The profile shown with long-dashed lines and plotted over the Fe $\lambda$ 1608 \AA \ profile has been estimated by assuming a solar (Fe/S) ratio and the S abundance determined for the 540 \kms\ component, see text.  \label{1}}

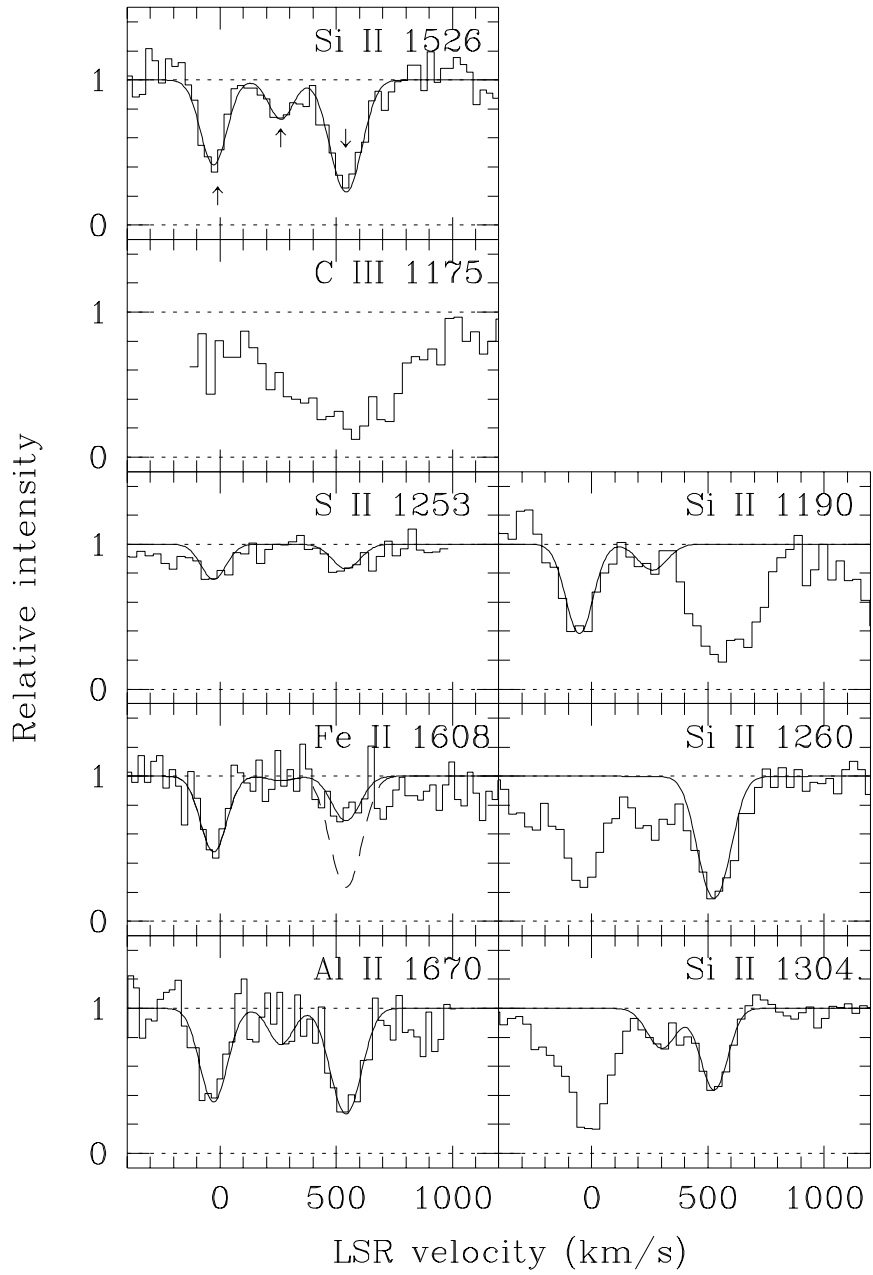
\end{document}